\documentclass[twocolumn,showpacs,preprintnumbers,amsmath,amssymb]{revtex4}
\usepackage{color}
\usepackage{graphicx,epsfig}
\usepackage[backref]{hyperref} 

\begin{document}
\title{High-Activity Perturbation Expansion for the Hard Square Lattice Gas} 
\author{Kabir Ramola and Deepak Dhar}
\affiliation{Department of Theoretical Physics, Tata Institute of Fundamental Research, Mumbai 400 005, India}
\date{\today}

\begin{abstract} 
We study a system of particles with nearest and next-nearest-neighbour exclusion on the square lattice (hard squares). This system undergoes a transition from a fluid phase at low density to a columnar ordered phase at high density. We develop a systematic high-activity perturbation expansion for the free energy per site about a state with perfect columnar order. We show that the different terms of the series can be regrouped to get a Mayer-like series for a polydisperse system of interacting vertical rods in which the $n$-th term is of order $z^{-(n+1)/2}$, where $z$ is the fugacity associated with each particle. We sum this series to get the exact expansion to order $1/z^{3/2}$.
\end{abstract}

\maketitle

\section{Introduction} 
Properties of systems with only hard core repulsive interactions between particles have been of continued interest in statistical physics. The temperature of the system plays no role and the phase transitions in these systems as a function of density are purely the result of geometrical effects of excluded volume interactions. These have been called geometrical phase transitions, the best studied of which are the percolation transition \cite{stauffer} and phase transitions in assemblies of hard spheres \cite{alder}. Many systems with different shapes of particles have been studied in the literature, for example squares \cite{pearce}, rods \cite{anandamohan}, triangles \cite{verberkmoes} and L-shaped molecules \cite{barnes}. The computation of the partition function of particles with finite exclusion volumes on a lattice remains an outstanding problem \cite{runnels_combs, baxter}. The model of hard hexagons on the triangular lattice and related models are the only known exactly soluble systems of this kind \cite{baxter_hardhexagon, difrancesco}. Such models in $d$-dimensions can also be related to directed and undirected lattice animal enumeration problems in $(d+1)$ and $(d+2)$-dimensions respectively\cite{dhar_animals, brydges}. 

\begin{figure}
\includegraphics[width=.45\columnwidth,angle=0]{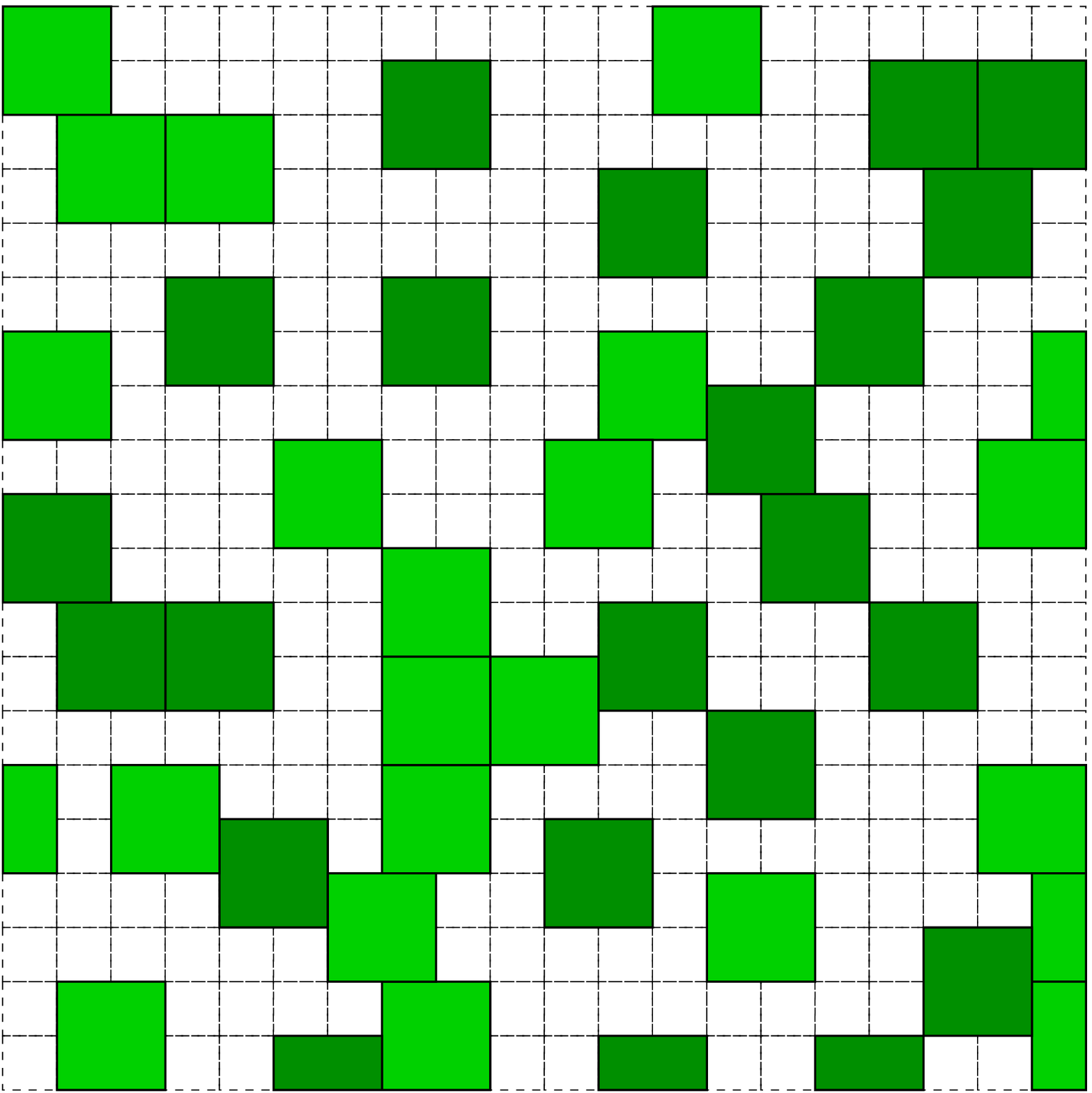}
\includegraphics[width=.45\columnwidth,angle=0]{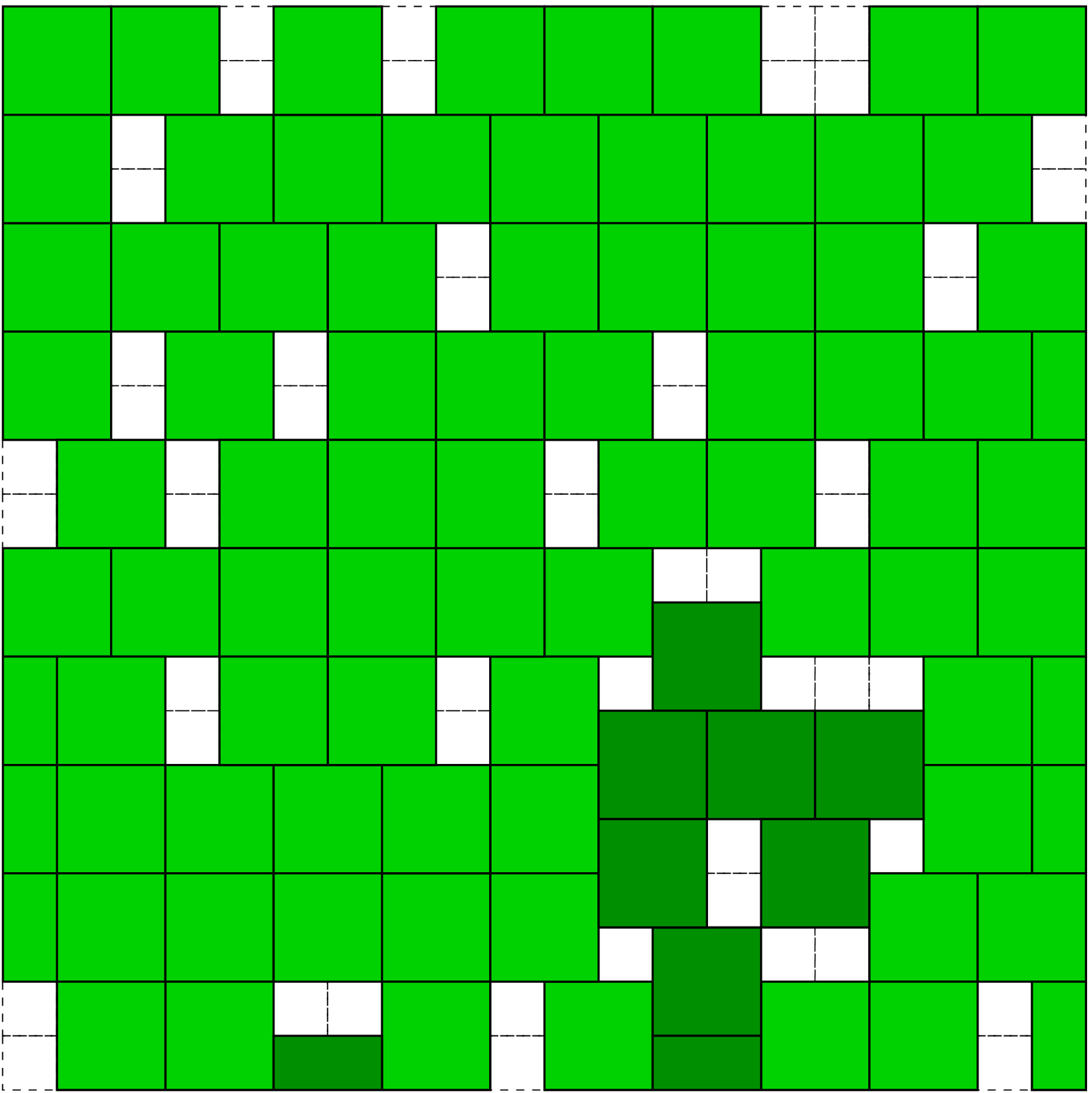}
\caption{A typical configuration of hard squares on the lattice at (left) low fugacity and (right) high fugacity. The light green squares correspond to particles on row $A$, whereas the dark green squares correspond to particles on row $B$. At high densities, the system is in a columnar ordered phase.}
\label{typical_figure}
\end{figure}

In many cases, the high density states exhibit crystalline order: the particles preferentially occupy a sublattice of the full lattice (for example hard hexagons on the triangular lattice and the nearest-neighbour-exclusion lattice gas on the square lattice). In such cases, it is straightforward to develop perturbation expansions in inverse powers of fugacity for thermodynamic quantities \cite{gaunt}. In this paper we study particles with nearest and next-nearest-neighbour exclusion on the square lattice (hard squares). Equivalently, each particle is a $2\times 2$ square that occupies $4$ elementary plaquettes of the square lattice (Fig. \ref{typical_figure}). This system displays ``columnar order'' at high densities where one of the odd or even rows (or columns) is preferentially occupied. In this case, the standard high-density expansion in inverse powers of the fugacity $z$ breaks down. It was realised quite early that the leading order correction to the high-activity expansion is of order $1/\sqrt{z}$ \cite{bellemans}, but a systematic expansion has not been developed so far. 

There have been only a few theoretical studies of the columnar ordered state so far, and the present understanding is not very satisfactory. Correlations in the columnar ordered state are hard to capture using mean-field like descriptions. For example, all the well-known approximation schemes like the mean-field theory and cluster variational approximations underestimate the value of the critical point $z_c$ for the hard squares problem by an order of magnitude.

In this paper we develop a systematic expansion of the free energy in the columnar ordered state extending the earlier treatment in \cite{bellemans}. This is a singular perturbation series in powers of $1/\sqrt{z}$. One introduces explicit symmetry breaking by ascribing different fugacities $z_A$ and $z_B$ to the particles on even and odd rows. The point $z_B = 0$ corresponds to a fully columnar ordered configuration. The perturbation expansion about the ordered state in powers of $z_B$ is a standard Mayer-like series \cite{mccoy}. For the case $z_A = z_B = z$, we show that the terms of the series can be regrouped and the resulting series can be thought of as a Mayer-like expansion of extended objects (vertical rods), but of arbitrary size. The term of order $1/z^{\frac{n+1}{2}}$ involves at most $n$ such objects and gets contributions from all possible sizes of these objects. We evaluate explicitly the contribution of terms corresponding to two rods, which gives exact results up to order $1/z^{3/2}$, adding one more term to the earlier known result of order $1/z$ \cite{bellemans}.

\section{The model}

We consider a system of particles with nearest and next-nearest-neighbour exclusion on the square lattice. This model has been the subject of many studies. Transfer matrix techniques indicate the existence of a phase transition in this model \cite{bellemans_2, ree_chestnut, nisbet_farquhar, domany, kinzel_schick}. Variational, density functional methods and virial expansions have also been used to study this problem \cite{ree_chestnut, aksenenko, lafuente_cuesta}. These studies indicate that at high densities the system is not sublattice ordered but exhibits columnar order. However, there is as yet no rigorous proof of this. In this context it seems worthwhile to develop systematic series expansions for the high-density phase, and investigate their convergence properties. Recent Monte Carlo evidence suggests that the transition from fluid to columnar order in this system is of second order, with exponents close to the two dimensional Ising model \cite{fernandes, nienhuis, ramola}.

\begin{figure}
\includegraphics[width=.4\columnwidth,angle=0]{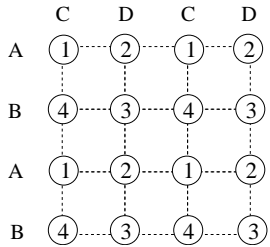}
\caption{We define four sublattices $1$ through $4$ on the square lattice. Rows containing the sites of sublattice $1$ and $2$ ($3$ and $4$) will be called $A$ ($B$) rows. Similarly, columns with sublattice sites $1$ and $4$ ($2$ and $3$) will be called $C$ ($D$) columns.}
\label{sublattice}
\end{figure}

We write the grand canonical partition function of the hard square lattice gas model as

\begin{equation}
\Omega(z) = \sum_{C} z^{n(C)},
\label{partition1}
\end{equation}
where $z$ is the fugacity of the particles, $n(C)$ is the number of particles in the configuration $C$ and the sum is over all allowed configurations of hard squares on the lattice.

The Landau free energy per site $f(z)$ is defined as the large-$N$ limit of $\left[-\log \Omega(z)\right]/N$, where $N$ is the total number of sites in the lattice. The density of particles at a particular fugacity is given by $ \rho(z) = -z \frac{\textmd{d}}{\textmd{d}z} f(z)$.

The low-activity Mayer series of this model can be computed using standard techniques \cite{mccoy}. We get \cite{hoover_alder}

\begin{equation}
 -f(z) = z - \frac{9}{2!} z^{2} + \frac{194}{3!} z^{3} -\frac{6798}{4!} z^{4} + \hdots. 
\end{equation}

This expansion has a finite radius of convergence determined by the singularity closest to the origin at $z_0 \approx -0.1$. The low-density fluid phase has short-ranged correlations between particles, but the high-density columnar-ordered phase has long-range order. Monte Carlo simulations yield the best estimates of the critical point as $z_c = 97.5 \pm 0.5$ \cite{fernandes,nienhuis}. The nature of the transition from the low density disordered state to the columnar ordered state at high densities has been the subject of several studies. The hard square lattice gas also arises as a limiting case of spin models with interactions up to the second nearest neighbour on the square lattice \cite{binder_landau, amar, zhitomirsky}. In the Landau theory paradigm, the four columnar ordered states in this system give rise to an XY model with four fold anisotropy \cite{domany_schick}. Such models are known to display non-universal behaviour with critical exponents that are governed by marginal operators, and naturally fall into the class of Ashkin-Teller-Potts models that exhibit such behaviour \cite{kadanoff_ashkin,jose_kadanoff,kadanoff_fourfold}. However, the critical exponents of this model have been hard to pin down from Monte Carlo simulations on small system sizes \cite{kinzel_schick}. 

There is some controversy about the nature of the transition from fluid to columnar order in this system. Some recent Monte Carlo studies suggested an Ising universality class \cite{fernandes}, however, this does not seem likely \cite{zhitomirsky,nienhuis}. Our own Monte Carlo studies and theoretical arguments also imply that the model is {\it not} in the Ising universality class \cite{ramola}.

\section{Expansion about the Crystalline Ordered State}

We consider the square lattice to be made up of four sublattices labelled $1$ to $4$ (Fig. \ref{sublattice}). Rows containing the sites of sublattice $1$ and $2$ are called $A$-rows, and those containing sites of sublattice $3$ and $4$ are $B$-rows. Similarly, columns with sublattice sites $1$ and $4$ are called C-columns and those with sublattice sites $2$ and $3$ are D-columns. We will specify the position of the square by the position of the top left corner of the square.

At the point $z = \infty$ the problem of computing the partition function reduces to one of calculating the number of perfect tilings of the lattice by hard squares. On a lattice of size $L$ X $L$ (where $L$ is assumed even), with open boundary conditions, clearly

\begin{equation}
\Omega_{\textmd{open}}(z \to \infty) = z^{L^2/4}.
\end{equation}

For a lattice with periodic boundary conditions in both $x$- and $y$- directions, it is easy to see that we have $\Omega_{\textmd{periodic}} = 4 (2^{L/2} - 1)z^{L^2/4}$.

For $z$ finite, there is a finite density of plaquettes not covered by squares in the system. Suppose we start with a fully packed configuration of squares, with all sites on sublattice $1$ and introduce a single vacancy. By sliding squares in the corresponding row, we can break the vacancy into two half-vacancies with in-between squares on the $2$-sublattice. These two vacancies can be moved arbitrarily far apart. Thus there are ${\cal O}( L^2)$ configurations of half-vacancies on each row, and there are $L/2$ possible rows. There is an equal number of configurations with half-vacancies along columns. This leads to 

\begin{equation}
\Omega_{\textmd{Open}}(z) = z^{L^2/4} \left(1 + \frac{1}{z} \mathcal{O}\left(L^3\right) + ...\right).
\end{equation}

This phenomena, which may be called deconfinement of half-vacancies, is the reason behind the failure of the standard Mayer expansion technique, in which the order $1/z$ term has a coefficient of the order of the volume of the system. Because the standard cumulant expansion fails, this series must be treated using different techniques.
 
A simple qualitative picture of the system with a small density of defects introduced on the background with perfect crystalline order is this: The pairs of half-vacancies may be pictured as joined by straight rods, which may be horizontal or vertical. The state with orientational (i.e. nematic) ordering of these rods corresponds to a state of the hard squares with columnar order. In fact, there is an exactly solved model of hard rods of variable length due to Ioffe {\it et al.} \cite{ioffe}, in which the activity of a rod is independent of its length, which shows a phase transition at a finite value of the activity of the rods. This phase transition is in the Ising universality class. However, we note that the precise weights in Ioffe {\it et al.}'s model are different from the hard squares case, and, in fact, the transitions are not expected to be in the same universality class.

\section{Expansion about the Columnar Ordered State}

In the columnar ordered state the system preferentially occupies one of the rows ($A$ or $B$) or columns ($C$ or $D$). In the $A$-ordered phase we have $(\rho_1 = \rho_2) > (\rho_3 = \rho_4$), where $\rho_i$ denotes the density of particles corresponding to the $i$th sublattice. The $B$, $C$, and $D$ phases are defined similarly. 

To quantify the nature of ordering in this system we define the following order parameters. The row order parameter of the system is defined to be

\begin{equation}
 O_{R} = 4 [(\rho_1 + \rho_2) - (\rho_3 + \rho_4)],
\label{row_OP}
\end{equation}
and the column order parameter is
\begin{equation}
 O_{C} = 4 [(\rho_1 + \rho_4) - (\rho_2 + \rho_3)].
\label{column_OP}
\end{equation}

Equivalently, we can also define a single $\mathbb{Z}_4$ complex order parameter 

\begin{equation}
O_{Z_4} = 4 \sqrt{2} [(\rho_1 - \rho_3) + i (\rho_2 - \rho_4)].
\end{equation}

The factor $4 \sqrt{2}$ has been introduced to make the maximum value of the order parameter $1$. The phase of the complex order parameter $O_{Z_4}$ takes the values $\pi /4, -3 \pi /4,-\pi /4$ and $3 \pi /4$ in the $A$, $B$, $C$, and $D$ phases respectively.

We now develop a high-activity perturbation series about the row-ordered state. We effectively integrate out the horizontal rods, and this generates a longer-ranged effective interaction between the vertical rods, in addition to the excluded volume interactions. 

As in \cite{bellemans}, we associate different fugacities to the particles on the $A$ and $B$-rows. The fugacities of the particles on row $A$ is $z_A$ and on row $B$ is $z_B$. $\Omega(z_A,z_B)$ is the partition function of the system with this explicit symmetry breaking between the two rows. $\Omega(z_A,0)$ corresponds to the fully columnar ordered state. We then expand the partition function of the system about a state with perfect columnar order in the $A$-phase. 

We write the partition function of the system as a formal expansion in terms of the fugacities of the particles on the $B$-rows (defects) and the corresponding partition functions of the $A$-rows. The partition function expansion about the columnar ordered state is

\begin{equation}
\frac{\Omega(z_A,z_B)}{\Omega(z_A,0)} = 1 + z_B W_1(z_A) + \frac{{z_B}^2}{2!} W_2(z_A) + \hdots.
\label{partition_expansion}
\end{equation}

Now, taking the logarithm of Eq. (\ref{partition_expansion}), we arrive at the cumulant expansion

\begin{equation}
\frac{1}{N} \log \Omega(z_A,z_B) = \kappa_0(z_A) + z_B \kappa_1(z_A) + \frac{{z_B}^2}{2!} \kappa_2(z_A) + \hdots.
\label{z_B_series}
\end{equation}

The calculation of $W_n(z_A)$ involves fixing a configuration ${\mathcal B}_n$ of $n$ particles on the $B$-sites. The weight of this configuration is defined to be $z_B^n \text{Prob}({\mathcal B}_n)$, where $\text{Prob}({\mathcal B}_n)$ is the probability that in the reference system with only $A$-particles, no site excluded by these $B$-particles will be occupied. We then sum these weights over all configurations ${\mathcal B}_n$. The negative of the logarithm of $\text{Prob}({\mathcal B}_n)$ is defined to be the effective interaction energy between the $B$-particles when the $A$-particles are integrated out.

It is straightforward to evaluate the first few terms in this expansion. When there are no $B$-particles in the lattice, the partition function of the system breaks up into a product of 1-d partition functions of particles on the $A$-rows (particles on different $A$-rows do not interact). These $A$-particles thus behave as a 1-d lattice gas with nearest neighbour exclusion. Thus we have

\begin{equation}
\Omega(z_A, 0) = \left[\Omega_{1d,L}(z_A)\right]^{L/2},
\label{1D_partitions}
\end{equation}
where $\Omega_{1d,L}(z_A)$ is the 1-d partition function of particles with nearest neighbour exclusion on a periodic ring of length $L$. This is easily seen to be

\begin{equation}
\Omega_{1d,L}(z_A) = {\lambda_{+}}^L + {\lambda_{-}}^L,
\label{periodic_omega}
\end{equation} 
where $\lambda_{\pm}$ are the eigenvalues of the 2 X 2 transfer matrix 

\begin{align}
\nonumber
&{\bf T} = \left[ \begin{array}{cc}1 & 1 \\ z_A & 0 \\ \end{array} \right],\\
\nonumber
&\hspace{1cm} \textmd{with}\\
&\lambda_{\pm} = \frac{1 \pm \sqrt{ 1 + 4 z_A}}{2}.
\label{eq:rho}
\end{align}

In the limit of large $L$, this gives us

\begin{equation}
\kappa_0( z_A) = \frac{1}{2} \log \lambda_+.
\end{equation}

At the next order, $W_1(z_A)$ involves fixing the position of a single particle on a $B$-site, say at $(x,y)$. Then the sum over $A$-particle configurations is restricted to those in which the sites $(x \pm1, y \pm 1)$ and $(x, y \pm 1)$ are not occupied. The partition function of a single row with this constraint is the partition function of a system of particles on an open chain of length $L-3$, and is easily calculated. Let $f_{000}$ be the probability that the sites in a randomly picked interval of length three on an $A$-row are all empty, in a 1-dimensional lattice gas with nearest neighbor exclusion at activity $z_A$. Then, 

\begin{equation}
W_1(z_A) = \frac{N}{2} \left(f_{000}\right)^2. 
\end{equation}

The expression for $f_{000}$ is easy to derive using the properties of the 1-d nearest-neighbor exclusion lattice gas. We have $f_{000} = \frac{1}{z_A} f_{010}$, where $f_{010} $ is the probability that a randomly picked site in the gas will have the occupation numbers $010$ at the three consecutive sites. Clearly, $f_{010}= \rho_{1d} (z_A)$, the density of the gas. Using the Eq. (\ref{eq:rho}), we get 

\begin{equation}
\rho_{1d} (z_A) = \frac{1}{2} - \frac{1}{2 \sqrt{1 + 4 z_A}}. 
\end{equation}

Therefore we obtain

\begin{equation}
\kappa_1(z_A) = \frac{1}{2} \left(f_{000}\right)^2 = \frac{1}{2} \left[\frac{\rho_{1d} (z_A)}{z_A}\right]^2.
\end{equation}

Therefore the leading contribution to the cumulant expansion from the single particle term is of order $\mathcal{O}\left(\frac{z_B}{{z_A}^2}\right)$.

The calculation of higher order terms $W_n(z_A)$ for $n \ge 2$ is similar. However, working order by order in $n$ is not very effective. In the series given in Eq. (\ref{z_B_series}), the series is in powers of $z_B$, with coefficients that are functions of $z_A$. Eventually, We would like to put $z_A = z_B =z$, and expand the series in inverse powers of $1/z$. Unfortunately, the leading behaviour of $W_n(z_A)$ for large $z_A$ is only $z_A^{-(n+1)}$. Then, arbitrarily large orders in $n$ are required even to get the correct result to order $1/z$ \cite{bellemans}.

This may be seen as follows: Consider the configuration of $n$ $B$-particles vertically above each other. It is easy to see that the probability that such a configuration would be allowed in the unperturbed ensemble is $[f_{000}]^{n+1}$, and hence is ${\cal O}\left(z_A^{-(n+1)}\right)$. Hence the contribution of this term to the perturbation expansion is of order ${z_B}^n z_A^{-(n+1)}$. For $z_B= z_A = z$ this is ${\cal O}(1/z)$ for all $n$.

It is easy to check that all configurations except these vertical rod-like configurations of $B$-particles do not contribute to order $1/z$. Hence, if we want to sum over all terms to order $1/z$, we group these configurations together, and identify them as a single vertical rod. A general configuration of $B$-particles would then be a group of non-overlapping vertical rods. In any configuration of $B$-particles, we define a rod containing a given occupied $B$-site $\vec{r}$ to be the set of all the consecutively occupied $B$-sites in the same column reachable from $\vec{r}$ using vertical steps of length $2$. Clearly, any configuration of $B$-sites has a unique description as a set of non-overlapping vertical rods. To avoid over-counting, two rods cannot sit directly on top of each other.

\section{High-Activity Expansion in terms of Rods}

We now develop an expansion in terms of the number of rod defects. We attach an additional activity factor $\epsilon$ to each rod, and now rewrite the summation in Eq.(\ref{partition_expansion}) as a sum over configurations involving different numbers of rods

\begin{equation}
\frac{\Omega(z_A, z_B)}{\Omega(z_A, 0)} = 1 + \epsilon R_1(z_A,z_B) + \epsilon^2 R_2(z_A,z_B) + \hdots,
\end{equation} 
where $R_n(z_A,z_B)$ denotes the contribution of the configurations with exactly $n$ rods to the partition function expansion. Now taking the logarithm, we arrive at the free energy expansion
\begin{align}
\nonumber
&\mathcal{F}(z_A,z_B) = - \frac{1}{N} \log \left( {\Omega(z_A, z_B)} \right), 
\\
&=F_{0}(z_A,0) + \epsilon F_1(z_A,z_B) + \epsilon^2 F_2(z_A,z_B) + \hdots,
\label{rod_cumulant}
\end{align} 
where $\mathcal{F}(z_A,z_B)$ denotes the free energy per site of the hard square lattice gas and

\begin{eqnarray}
\nonumber
F_1(z_A,z_B) &=& - \frac{1}{N} R_1(z_A,z_B),\\
\nonumber
F_2(z_A,z_B) &=& - \frac{1}{N} \left( R_2(z_A,z_B) - \frac{{R_1(z_A,z_B)}^{2}}{2} \right).\\
&\vdots&
\end{eqnarray}

We evaluate the free energy expansion formally in powers of $\epsilon$. At the end of the calculation, we set $\epsilon =1$. $F_{0}$ denotes the contribution from the term when there are no rods in the system. From Eq. (\ref{rod_cumulant}) we have

\begin{equation}
F_{0} = -\kappa_0 (z_A).
\end{equation}

\begin{figure}[ht]
\includegraphics[width=.5\columnwidth]{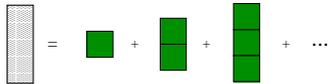}
\caption{The configurations contributing to a single rod term}
\label{single_rods}
\end{figure}

Now consider configurations with exactly one rod. The weight of a rod with length $n$ is easily seen to be ${f_{000}}^{n+1} {z_B}^{n}$. There are $N_B= N/2$ possible positions for each size $n$ of the rod. Summing over all possible values of $n$, we get 

\begin{equation}
F_1 = -\frac{N_B}{N} \sum_{n = 1}^{\infty} (f_{000})^{n+1} {z_{B}}^n,
\end{equation}
which yields,
\begin{equation}
F_{1} = -\frac{1}{2} \left(\frac{z_{B} {f_{000}}^2}{1 - z_{B} f_{000}} \right).
\end{equation}

It is convenient to define a parameter $\gamma = z_B f_{000}$. Then we have

\begin{equation}
F_{1} = -\frac{\gamma f_{000}}{2(1 - \gamma)}.
\end{equation}

We next evaluate the two rod term, with

\begin{equation}
F_2 = -\sum_{\vec{r}_1,l_1} \sum_{\vec{r}_2,l_2} \left[ w(\vec{r}_1,l_1;\vec{r}_2,l_2) - \frac{w(\vec{r}_1,l_1) w(\vec{r}_2,l_2)}{2} \right].
\end{equation}
where $w(\vec{r}_1,l_1;\vec{r}_2,l_2)$ denotes the weight of the two-rod configuration with the centre of the first rod of length $l_1$ at position $\vec{r}_1$ and the centre of the second rod of length $l_2$ at position $\vec{r}_2$, in the partition function expansion. $w(\vec{r}_1,l_1)$ denotes the weight of a single rod configuration. 

\begin{figure}[ht]
\includegraphics[width=.9\columnwidth]{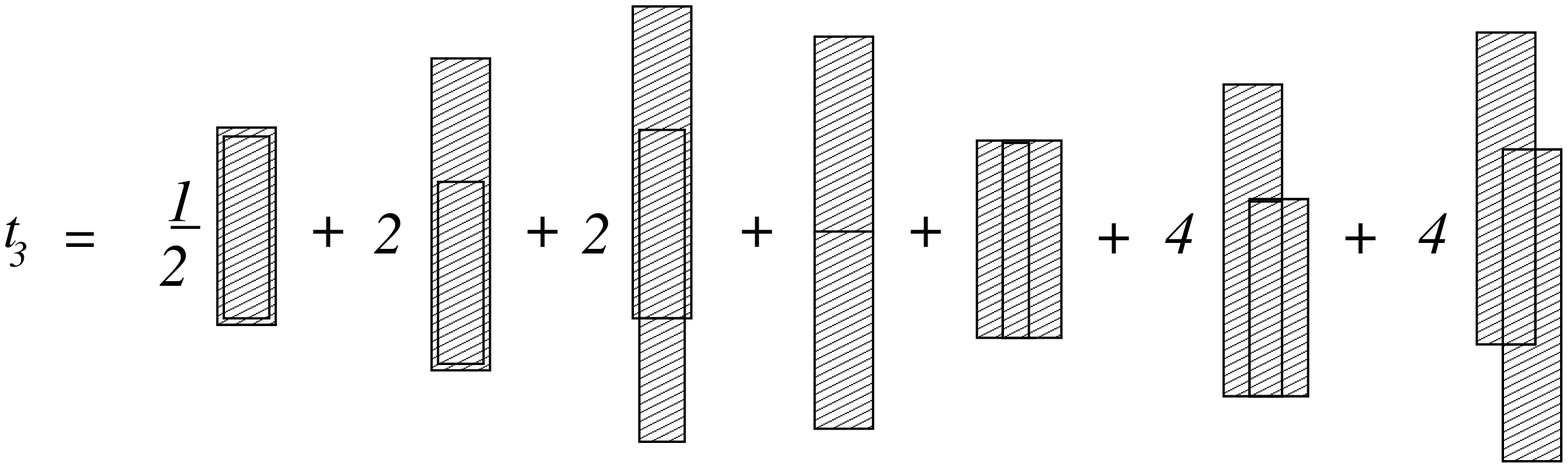}
\caption{The configurations contributing to the overlapping rods term}
\label{intersecting_configs}
\end{figure}

In this summation, the contribution from the two-rod terms that have zero interaction are exactly cancelled in the corresponding cumulant expansion. The configurations with non-zero interaction between the rods are ones where at least one row is touched by both the rods. These configurations can be classified as follows:\\
1) Adjacent rods (configurations in which only one row is touched by the ends of two rods from opposite sides (Fig. \ref{adjacent_rods})),\\
2) Two rods which share a non-zero interval in the Y-direction. In this case there are a finite number of rows occupied by both rods (Fig. \ref{overlapping_configs}),\\
3) Overlapping rods, where a finite area of the lattice is occupied by both rods (The weight of the configuration $w(\vec{r}_1,l_1;\vec{r}_2,l_2)$ is zero, but $w(\vec{r}_1,l_1)w(\vec{r}_2,l_2)$ is finite)(Fig. \ref{intersecting_configs}).\\

We deal with these three terms separately. We have 

\begin{equation}
 F_2 = t_1 + t_2 + t_3.
\end{equation}

\subsubsection{Adjacent rods}

\begin{figure}[ht]
\includegraphics[width=.3\columnwidth]{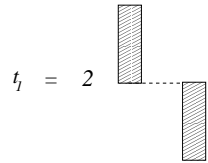}
\caption{The configurations contributing to the adjacent rods term}
\label{adjacent_rods}
\end{figure}

In the calculation of the term $t_1$, we deal with all configurations in which two rods touch an $A$-row from different sides. Then for this row, we need to calculate the correlation function in the reference problem that two triplets of three consecutive sites at a given distance are both empty. This is easily expressed in terms of the correlation function of occupied sites in the reference system. Let the pair correlation function at a separation $\Delta$ between the centres of the triplets in the 1-d nearest neighbour exclusion gas be $G(\Delta)$. Then it is easily seen that 

\begin{eqnarray}
\nonumber
G\left(\Delta \right) &=& \left( f_{000}\right) ~~~\text{for $\Delta = 0$},\\
\nonumber
&=& \left( f_{000}\right) \frac{1}{\lambda_+} ~~~\text{for $|\Delta| = 1$},\\
\nonumber
&=& \left(f_{000}\right)^2 \left(1- \alpha^{|\Delta|-1}\right) ~~~\text{for $|\Delta| \ge 2$},\\
\label{Wweights} 
\end{eqnarray}
where
\begin{equation}
\alpha = \frac{\lambda_{-}}{\lambda_{+}} = -1 + \frac{1}{{z_A}^{1/2}} - \frac{1}{2} \left(\frac{1}{z_A}\right) + \mathcal{O}\left( \frac{1}{{z_A}^{3/2}}\right).
\end{equation}

One of the rods extends upwards with length $n_a$ and the other rod extends downwards with a length $n_b$ (Fig. \ref{adjacent_rods}), the factor of $2$ accounts for the symmetry related diagrams. Summing over all configurations and subtracting the disconnected part, we have

\begin{align}
\nonumber
&t_{1} = -\sum_{n_a=1}^{\infty} \sum_{n_b=1}^{\infty} \gamma^{n_a + n_b} \\
&\times \left( \left(\frac{f_{000}}{\lambda_{+}} - {f_{000}}^2 \right) + \sum_{\Delta \ge 2}^{\infty} {f_{000}}^2 \left[\left(1 - \alpha^{|\Delta|-1}\right) -1 \right]\right).
\end{align}

This series can be easily summed. We have

\begin{equation}
t_{1} = -\frac{\gamma^2 f_{000}}{(1 - \gamma)^2} \left(\frac{1}{\lambda_{+}} - \frac{f_{000}}{1 - \alpha} \right).
\end{equation}

\subsubsection{Rods which share a non-zero interval in the Y-direction}

\begin{figure}[ht]
\includegraphics[width=.6\columnwidth]{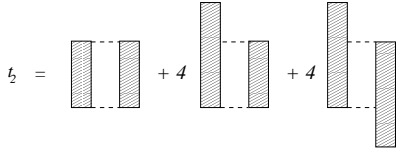}
\caption{The configurations contributing to the term involving rods which share a finite Y-interval. The factors multiplying the diagrams account for the symmetry related configurations.}
\label{overlapping_configs}
\end{figure}

In the evaluation of the term $t_2$, we have two rods with $n_o$ $B$-particles sharing a common interval in the Y-direction. Now, one of the rods can have a part extending above this section (Fig. \ref{overlapping_configs}), this can happen in $4$ ways. Also, the two rods can both have a finite extension above this section, this can occur in two distinct ways as follows: one of the rods extends both above and below the other, or one rod extends above and the other below. Both these cases yield the same weights. The extending sections above and below the overlap are of lengths $n_a$ and $n_b$. We have 
\begin{align}
\nonumber
&t_{2} = -\frac{1}{2} \left( 1 + 4 \sum_{n_{a} = 1} ^{\infty} \gamma^{n_{a}} + 4 
\sum_{n_{a} = 1} ^{\infty} \sum_{n_{b} = 1}^{\infty} \gamma^{n_{a}+ n_{b}}\right)\\
&\times \left(\sum_{n_{o} = 1}^{\infty} \sum_{\Delta \ge 2}^{\infty} \gamma^{2 n_{o} } (f_{000})^2 \left[\left(1 - \alpha^{|\Delta|-1}\right)^{n_{o} + 1} -1 \right]\right).
\label{term2}
\end{align}
The sum within the first brackets can be performed easily to give $\left( \frac{1 + \gamma}{1- \gamma} \right)^2$. 
The summation in the second brackets can be evaluated and we have
\begin{align}
\nonumber
&t_{2} = -\frac{1}{2} \left( \frac{1 + \gamma}{1- \gamma} \right)^2 \gamma^2 (f_{000})^2\\ 
&\times \left( \sum_{\Delta \ge 2}^{\infty} \frac{ \left(1 - \alpha^{|\Delta|-1}\right)^{2}}{1- \gamma^2 \left(1 - \alpha^{|\Delta|-1}\right)} - \frac{\left( f_{000} \right)^{2} \gamma^2 }{1- \gamma^2 }\right).
\end{align}
We can rewrite this in a better form as
{\small
\begin{align}
&t_{2} = -\frac{1}{2} \left( \frac{1 + \gamma}{1- \gamma} \right)^2 (f_{000})^2 \left( \sum_{r = 1}^{\infty} \frac{\left(1 - \alpha^{r}\right)^{2}}{\gamma^{-2} - \left(1 - \alpha^{r}\right)} - \frac{1 }{\gamma^{-2} - 1}\right).
\end{align}
}
Now, since $\alpha$ is negative, the terms in the summand series are oscillatory. We thus split the terms into even and odd powers of $\alpha$. We write

\begin{align}
&t_{2} = -\frac{1}{2} \left( \frac{1 + \gamma}{1- \gamma} \right)^2 \left( f_{000} \right)^{2} \left( S_{odd} + S_{even} \right),
\end{align}
where
\begin{align}
&S_{even} = \left( \sum_{n = 1}^{\infty} \frac{\left(1 - \alpha^{2 n}\right)^{2}}{\gamma^{-2} - \left(1 - \alpha^{2 n}\right)} - \frac{1 }{\gamma^{-2} - 1}\right),
\end{align}
and
\begin{align}
S_{odd} = \left( \sum_{n = 0}^{\infty} \frac{\left(1 + \alpha^{2 n + 1}\right)^{2}}{\gamma^{-2} - \left(1 + \alpha^{2 n + 1}\right)} - \frac{1 }{\gamma^{-2} - 1}\right).
\end{align}

For large $z$, $\alpha \approx -1 + z^{-1/2}$, and the summand in $S_{odd}$ and $S_{even}$ is a slowly varying function of $n$. To evaluate these expressions we can approximate the sum over the discrete values by an integral over the appropriate range of $n$. The error is only of order $1/z$. We get 
\begin{align}
\nonumber
&S_{even} \simeq \int_{n = \frac{1}{2}}^{\infty} dn \left(\frac{\left(1 - \alpha^{2 n}\right)^{2}}{\gamma^{-2} - \left(1 - \alpha^{2 n}\right)} - \frac{1 }{\gamma^{-2} - 1} \right) =\\
&\frac{1}{-2 \log (-\alpha)}\left(-\alpha - \frac{\gamma^{-4}}{\gamma^{-2}-1} \log \left(\frac{\gamma^{-2} -1 + \alpha}{\gamma^{-2} - 1}\right)\right),
\end{align}
and 
\begin{align}
\nonumber
&S_{odd} \simeq \int_{n = - \frac{1}{2}}^{\infty} dn \left( \frac{\left(1 + \alpha^{2 n + 1}\right)^{2}}{\gamma^{-2} - \left(1 + \alpha^{2 n + 1}\right)} - \frac{1 }{\gamma^{-2} - 1}\right)\\
&= \frac{1}{2 \log (-\alpha)} \left(1 + \frac{\gamma^{-4}}{\gamma^{-2}-1} \log \left(\frac{\gamma^{-2} -2}{\gamma^{-2}-1}\right)\right).
\end{align}

\subsubsection{Overlapping Rods}

In the calculation of the overlapping rods term, we deal with all possible configurations in which two rods which partially, or fully overlap. In this case the two rods can be at a distance $\Delta = 0$ or $\Delta = \pm 1$ away from each other. The length of the overlap is $n_o$ and the parts extending above and below are of lengths $n_a$ and $n_b$ respectively. There is also a contribution from the forbidden configuration in which two rods sit directly on top of each other. We have

\begin{align}
\nonumber
&t_{3} = \frac{3}{4} \left( 1 + 4 \sum_{n_{a} = 1} ^{\infty} \gamma^{n_{a}} + 4 
\sum_{n_{a} = 1} ^{\infty} \sum_{n_{b} = 1}^{\infty} \gamma^{n_{a}+ n_{b}}\right)\\
& \times \left(\sum_{n_{o} = 1}^{\infty} (f_{000})^2 \gamma^{2 n_{o}} \right) +\frac{1}{2}\sum_{n_{a} = 1} ^{\infty} \sum_{n_{b} = 1}^{\infty}(f_{000})^2 \left( \gamma\right)^{n_{a}+ n_{b}},
\end{align}
which can be easily simplified to give
\begin{align}
&t_{3} = \frac{3}{4} \left( \frac{1 + \gamma}{1- \gamma} \right)^2 \left(\frac{{f_{000}}^2 \gamma^2}{1- \gamma^2}\right) + \frac{1}{2} \left(\frac{{f_{000}} \gamma}{1-\gamma}\right)^2.
\end{align}

\subsection{High-Activity Expansion for the Free Energy}

Setting $\epsilon = 1$ in Eq. (\ref{rod_cumulant}) we obtain a high-activity expansion for the free energy of the system. The term $F_1$, corresponding to single rods, has a leading contribution of order $1/z$. The term $F_2$, corresponding to two-rod configurations has a leading contribution of order $1/z^{3/2}$. The order $1/z^{3/2}$ contribution to the two-rod term comes from two sources, a) when two adjacent rods separated by a distance $1$ touch the same row from opposite sides and b) a summation over the distance between two rods which share a finite interval in the Y-direction. The sum over the distance in this case yields a factor of order $\sqrt{z}$ because the correlation length between the rods is $\xi = \log{\alpha} \sim \sqrt{z}$. The term of order $1/z^{2}$ gets contributions from configurations involving three rods, in addition to the above single rod and two rod terms. For $n$ rods, the summation over the distance between rods yields a leading order contribution of at most $z^{\frac{n-1}{2}}$. Hence in the evaluation of the term of order $z^{\frac{n+1}{2}}$, only terms involving $n$ rods need to be considered.

Using our expressions for $F_0$, $F_1$ and $F_2$, we can generate the exact series expansion for the free energy and the density of the hard square lattice gas up to order $1/z^{3/2}$. We have

\begin{align}
\nonumber
& -f(z) =\\
&\frac{1}{4} \log z + \frac{1}{4 z^{1/2}} + \frac{1}{4 z} + \frac{\left(3 \log{\left(\frac{9}{8}\right)} +\frac{11}{96}\right)}{z^{3/2}} + \mathcal{O}\left(\frac{1}{z^2}\right),\\
\nonumber
&\hspace{3cm} \textmd{and}\\
& \rho(z) = \frac{1}{4} - \frac{1}{8 z^{1/2}} - \frac{1}{4 z} - \frac{\left(\frac{9}{2} \log{\left(\frac{9}{8}\right)} +\frac{11}{64}\right)}{z^{3/2}} + \mathcal{O}\left(\frac{1}{z^2}\right).
\label{density_series}
\end{align}

Higher order terms can in principle be evaluated in a similar manner. The term of order $1/z^{2}$ gets contributions from configurations with up to three rods. This involves many more diagrams with more complicated summations, and will not be attempted here. However the convergence of this series for $z \gtrsim 150$ to the actual values seems quite good. For example at $z = 200$ and $150$ Eq. (\ref{density_series}) gives $\rho = 0.2396$ and $0.2377$, to be compared with results from Monte Carlo simulations which yield $\rho = 0.2395$ and $0.2374$ respectively.

\subsection{Order Parameter Expansion}

In this section we compute the high-activity expansion for the row order parameter which is defined in Eq. (\ref{row_OP}). We have

\begin{equation}
O_{R}(z_A, z_B) = 4 \left[z_A \frac{\partial}{\partial z_A} - z_B \frac{\partial}{\partial z_B}\right] \left(-\mathcal{F}(z_A, z_B)\right).
\end{equation}

Using the terms in the free energy expansion above we can generate the exact order parameter expansion to order $1/z^{3/2}$. We have
{\small
\begin{align}
\nonumber
&O_{R}(z,z) =\\
& 1 - \frac{1}{2 z^{1/2}} - \frac{5}{z} - \left( \frac{395}{16} + 50 \log \left(\frac{9}{8}\right) \right)\frac{1}{z^{3/2}} + \mathcal{O}\left(\frac{1}{z^2}\right).
\end{align}
}

Using the terms up to order $1/z^{3/2}$ in the order parameter expansion, we estimate the critical point of the system by solving for the point $O_{R} = 0$. We obtain an estimate of the critical point $z_c = 14.86$, which is significantly better than the estimate $z_c = 6.25$ obtained by using just the first three terms of the series.

\section{Discussion}

In this paper we have developed a perturbation expansion about the columnar ordered state for the hard square lattice gas. We identified the basic objects of excitations about the ordered state, namely vertical rods. We showed that only configurations with at most $n$ rods contribute up to ${\mathcal O}(z^{-(n+1)/2})$ in the expansion in inverse powers of $z^{-1/2}$, and we explicitly summed the contribution from the terms containing two rods that provides exact results up to order $1/z^{3/2}$. It is possible to extend the series expansion developed in this paper to other systems that display columnar ordered behaviour at high densities. In particular for the $k \times k$ hard square lattice gas a similar procedure can be used to generate the high-activity expansion with terms of powers of $z^{-1/k}$ appearing. Other systems displaying columnar ordering at high densities such as hard rectangles on the square lattice are also amenable to similar treatments. 

The method can be extended to three dimensional systems that exhibit columnar order. Consider a system of $ 2 \times 2 \times 2 $ hard cubes on the cubic lattice. Consider a state with perfect columnar order, with columns along the $z$-direction. Then its cross-section in the xy-plane would be a fully packed configuration of $2 \times 2$ hard squares. Then, as discussed above, there are approximately $ 2^{L/2}$ different fully-packed hard squares configurations. We can develop a perturbation theory in inverse powers of $\sqrt{z}$, about any of these states. However, it is easily seen that the first correction term (of order $1/z$) is different for different states, and is largest when all the columns are arranged in a periodic superlattice with square symmetry, and spacing $2$. These states then outweigh all other states, for nonzero $1/z$, in the thermodynamic limit. This is an example of ``order-by-disorder'' \cite{villain} in this system. The extended objects that contribute to order $1/z$ in the $z_A = z_B = z$ series in this case turn out to be rigid rods along the $x$- or $y$- directions. The interactions between the rods are similar to the two dimensional case, but the summations are harder to do in closed form. This seems to be an interesting direction for future studies. Another problem of interest is to establish lower bounds on the radius of convergence of these expansions.

\section*{Acknowledgments}
We acknowledge useful discussions with K. Damle. K.R. acknowledges productive discussions with R. Dandekar and P. Narayan. D.D. would like to acknowledge support from the Indian DST under the grant DST-SR/S2/JCB-24/2005.

\end{document}